\documentclass[manuscript]{aastex61}
\usepackage{graphicx}
\usepackage{xcolor}
\usepackage[utf8]{inputenc}
\usepackage[percent]{overpic}
\shorttitle{Fragmentation of Fullerenes to Linear Carbon Chains.}
\shortauthors{Strelnikov et al.}

\begin{document}
\title{Fragmentation of Fullerenes to Linear Carbon Chains.}

\correspondingauthor{Dmitry V. Strelnikov}
\email{dmitry.strelnikov@kit.edu}
\author{Dmitry V. Strelnikov}
\author{Manuel Link}
\author{Manfred M. Kappes}
\affiliation{Karlsruhe Institute of Technology (KIT), Division of Physical Chemistry of Microscopic Systems, Karlsruhe, Germany}

\begin{abstract}
Small cationic fullerene fragments, produced by electron impact ionization
of C$_{60}$, were mass-selected and accumulated in cryogenic Ne matrixes.
Optical absorption spectroscopy of these fragments with up to 18 carbon
atoms revealed linear structures. Considering the recent discovery
of fullerenes in Space and the very strong absorptions of long linear
carbon clusters both in the UV-Vis and IR spectral regions, these
systems are good candidates to be observed in Space. We present laboratory
data, supported by quantum-chemical calculations and discuss the relevance
of long carbon chains for astronomy. 
\end{abstract}
\keywords{methods: laboratory: molecular \textemdash{} ISM: lines
and bands \textemdash{} ISM: molecules}

\section{Introduction}

After the recent discovery of C$_{60}$, C$_{70}$ and C$_{60}^{+}$
in Space \citep{C60_space1,C60_space2,C60_space3,C60_space2013} and
in particular after the unequivocal attribution of several of the
diffuse interstellar bands (DIBs) to NIR absorption bands of C$_{60}^{+}$
\citep{MaierDIBs2015,MaierDIBs2015_2}, fullerenes have become a hot
topic in the astronomical community. Researchers are now also beginning
to consider other fullerene-derived or fullerene-related species,
which could be present and observable in Space \citep{Omont}. One
can divide these species into four classes: (i) reaction products
of fullerenes with abundant atoms, ions and molecules in Space, (ii)
fullerene fragmentation products, (iii) precursors of fullerenes,
and (iv) species, which could be formed together with fullerenes.
We have decided to explore the second class of species, concentrating
in particular on fullerene fragments expected to be strongly light
absorbing. Fullerenes are known to fragment predominantly via a sequential
C$_{2}$-loss cascade, leading to C$_{58}$, C$_{56}$, C$_{54}$
and smaller fullerenes down to C$_{32}$ \citep{Bekkerman2006}. Nevertheless,
it was already established in the early days of fullerene mass spectrometry
that given enough excitation energy, fullerene precusors could also
fragment into much smaller carbon cluster species \citep{Volpel1993}.

In the present work we investigate these smaller carbon clusters (C$_{n}^{+}$,
n$<$25), formed upon multifragmentation of fullerenes by high energy
electron impact ionization of sublimed C$_{60}$. Clusters in this
size range have been the object of studies for many years --
both from the point of view of cluster science and also from the vantage
of astronomy and diffuse interstellar bands. However, the species
in question have typically been generated using ``bottom-up'' approaches,
e.g, by aggregation of atomic (and small molecule) carbon vapour.
The corresponding spectroscopic studies imply that for all accessible
charge states (+/0/-) and for many nuclearities, ring and chain isomers
can be formed -- sometimes simultaneously. Nevertheless,
spectral coverage is often incomplete and assignment (to specific
structures and sizes) sometimes questionable. On the basis of ion
mobility spectrometry, it appears that carbon cluster aggregation
growth regimes can become kinetically constrained -- thus
giving rise to isomer distributions which are far from thermodynamic
equilibrium and which vary strongly with source conditions (\citet{Fromherz2002}
-- C$_{10-13}^{-}$ chains and rings; \citet{Koyasu2012}
-- C$_{7-10}^{+}$ rings and chains). The mass spectra
obtained upon multifragmenting fullerenes down to small cluster sizes
show abundance maxima and minima which are similar to those obtained
in aggregation growth \citep{Volpel1993,Cheng1996,Hunsche1996,Rohlfing1984}.
However, until now it has not been known which molecular structures
are formed and how these structures relate to those obtained by aggregation
growth. Our measurements described below indicate for the first time
that the molecular structures obtained by electron-impact multifragmentation
of C$_{60}$ include carbon chains. The state of knowledge concerning
the structures and spectroscopy of small carbon molecules C$_{n}$
(smaller than fullerenes) was thoroughly reviewed in 1998 \citep{VanOrden1998}.
Astronomical relevance of carbon clusters was recently reviewed \citep{Zack2014,MaierRev2017}.
Small carbon chain molecules such as C$_{2}$, C$_{3}$ and C$_{5}$
have already been unequivocally observed in circumstellar environments
\citep{Hrivnak1991,Hinkle1988,Bernath1989}. C$_{2}$ and C$_{3}$
were also detected further away from stars in the diffuse clouds \citep{Souza1977,Maier2001}.
Furthermore, linear cyanopolyynes as long as HC$_{9}$N \citep{Broten1978}
have been detected by radio astronomy. The idea that carbon chain
molecules may be responsible for the DIBs was initially proposed by
\citet{Douglas1977} and has been hotly debated over the years. Our
new findings provide additional support for carbon chain molecules
in Space.

\section{Experimental Details}

Our experimental setup is designed to study mass-selected, ionic species,
soft-landed and trapped in cryogenic matrixes \citep{Depo2C60}. The
apparatus allows investigation of such matrix isolated species by
optical absorption spectroscopy, Raman spectroscopy and Laser-Induced-Fluorescence
(LIF) measurements. One of the unique features of the setup is the
possibility to cover a very broad spectral range from UV to far-IR
without changing the sample, which is achieved by using diamond optics
\citep{DiamondBS}. In the current study cationic fullerene fragments,
produced by an electron-impact (EI) ionization source and also corresponding
neutral and anionic species, formed upon charge changing during deposition
into the matrix, were characterized by one-photon absorption spectroscopy.

In the preliminary experiments reported here, we deposited mass selected
C$_{15}^{+}$ and C$_{18}^{+}$ (together with the corresponding fullerene
multications, C$_{60}^{4+}$ and C$_{54}^{3+}$, respectively), into
neon matrixes. Despite the fact that the ion currents for these species
were quite low (0.3\textendash 1~nA), it turns out that UV-Vis and
IR spectra can be measured after a few hours of deposition. This means
that corresponding absorption cross sections must be very large.

\section{Results}

\begin{figure}
\centering \includegraphics[width=0.7\textwidth]{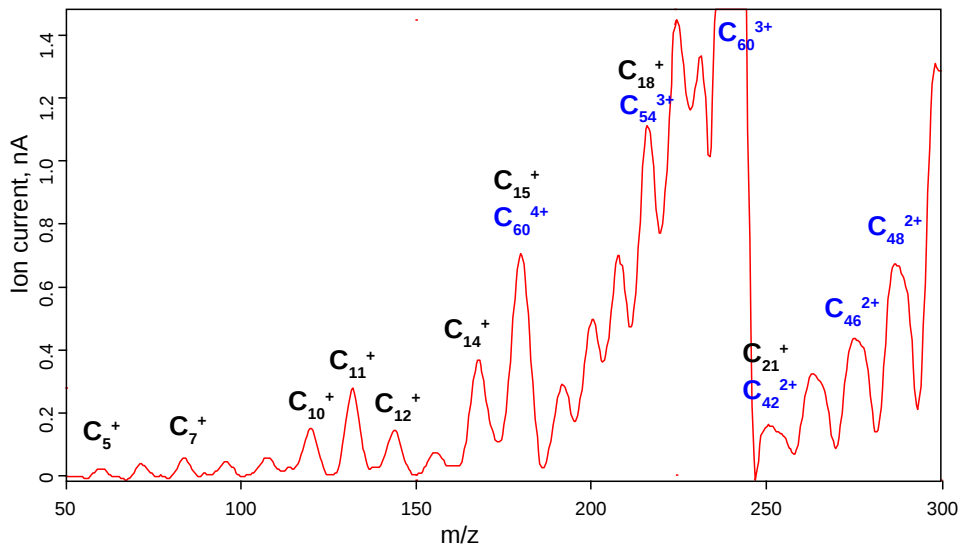}
\caption{\label{CnMS} Typical cationic mass spectrum of fullerene fragmentation
and ionization products by electrons at 250~eV. Note, that the mass
resolution of our experimental setup (optimized for high ion throughput)
does not allow differentiation of the isotopomers of C$_{15}^{+}$
and C$_{60}^{4+}$. Ion current for $m/z=240$ is 4.5~nA.}
\end{figure}

Fig. \ref{CnMS} shows a partial cation mass spectrum of fragment
ions obtained upon 250~eV electron impact ionization of sublimed
C$_{60}$. The spectrum resembles previously published mass spectra
of carbon clusters in that it manifests local ion abundance maxima
at m/z 132 and 180 amu \citep{Cheng1996,Hunsche1996,Volpel1993,Rohlfing1984}.
These mass peaks are thought to correspond primarily to C$_{11}^{+}$
and C$_{15}^{+}$ respectively. However, the underlying electron impact
fragmentation cascade proceeds in part via multiply charged fullerene
cages. Therefore, some of the signal in particular at m/z=180 is contributed
to by multiply charged fullerene cages, e.g. C$_{60}^{4+}$.

\subsection{Electronic Spectroscopy}

\begin{figure}
\centering \includegraphics[width=0.95\textwidth]{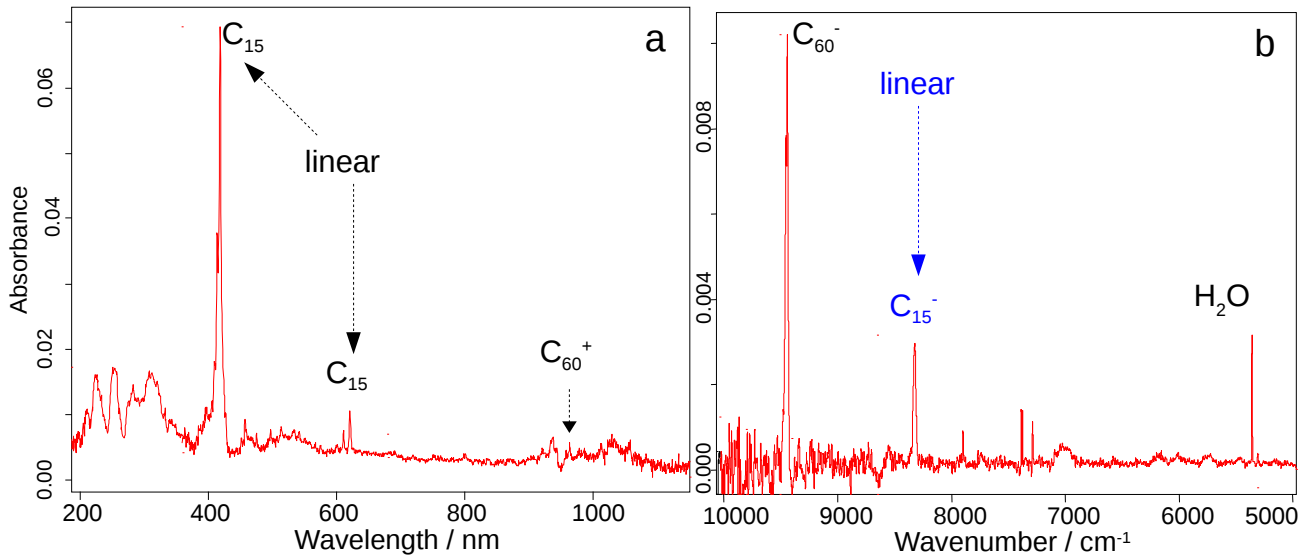}
\caption{\label{C15VISNIR} \textbf{a:} UV-NIR absorption spectrum of C$_{15}$
in Ne at 5K, obtained after deposition of ca. 1000~nAmin of C$_{15}^{+}$/C$_{60}^{4+}$
at about 100~eV kinetic energy. Bands between 200 and 350~nm may
be caused partly by optical interference. \textbf{b:} NIR absorption
spectrum of the same sample, obtained with an FTIR spectrometer.}
\end{figure}

C$_{15}^{+}$ (+ C$_{60}^{4+}$) deposition at an average kinetic
energy of about 100~eV resulted in relatively clean absorption spectra
(Fig.~\ref{C15VISNIR}). From previous measurements of the Maier
group (in which the clusters were formed in a Cs-sputtering ion source
and mass-selected or -- alterntively -- produced
by laser ablation of graphite without mass-selection), it is clear
that the absorptions observed in the UV-Vis and NIR ranges can be
assigned to linear isomers of C$_{15}$ and C$_{15}^{-}$ \citep{Forney1996,Wyss1999}.
At this point we cannot yet say where the absorptions of C$_{15}^{+}$
are, as this requires additional measurements in matrixes doped by
electron scavengers (to change the C$_{15}^{+/0/-}$ charge distribution).
However, TDDFT calculations predict the strongest absorption of C$_{15}^{+}$
to be in the vicinity of the absorption of neutral C$_{15}$, see
\ref{TC}. 

\begin{figure}
\centering \includegraphics[width=0.95\textwidth]{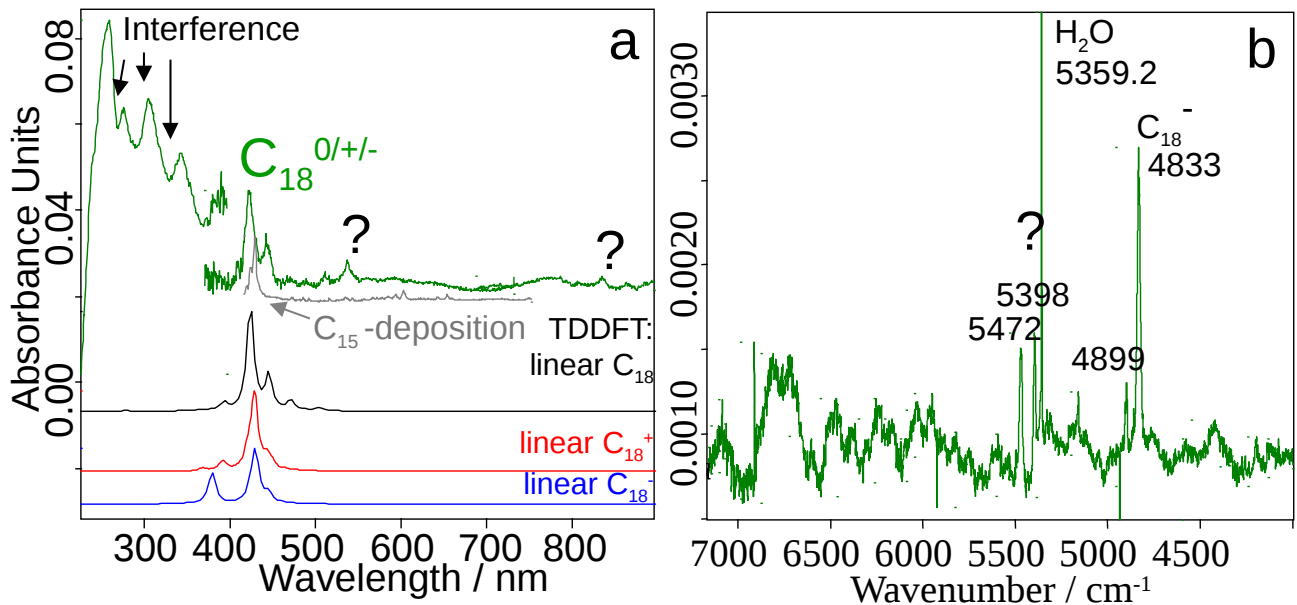}
\caption{\label{C18VISNIR}\textbf{ a:} UV-NIR absorption spectrum of C$_{18}$
in Ne at 5K (green), obtained after deposition of ca. 1800 nAmin of
C$_{18}^{+}$/C$_{54}^{3+}$ at about 120~eV kinetic energy. Bands
between 200 and 350~nm are caused by interferences. C$_{15}$ absorption
(gray) is shown as a reference. TDDFT calculations are shown without
wavelength scaling. \textbf{b:} NIR absorption spectrum of the same
sample, obtained with an FTIR spectrometer.}
\end{figure}

C$_{18}^{+}$ deposition resulted in the spectra presented in Fig.~\ref{C18VISNIR}.
Only NIR absorptions of linear C$_{18}^{-}$ in Ne were previously
reported -- in experiments without mass-selection, using
an extrapolation procedure based on measurements of shorter chains
\citep{Freivogel1995}. The UV-Vis spectrum of mass-selected C$_{18}$
is presented here for the first time.

\begin{figure}
\centering \includegraphics[width=0.95\textwidth]{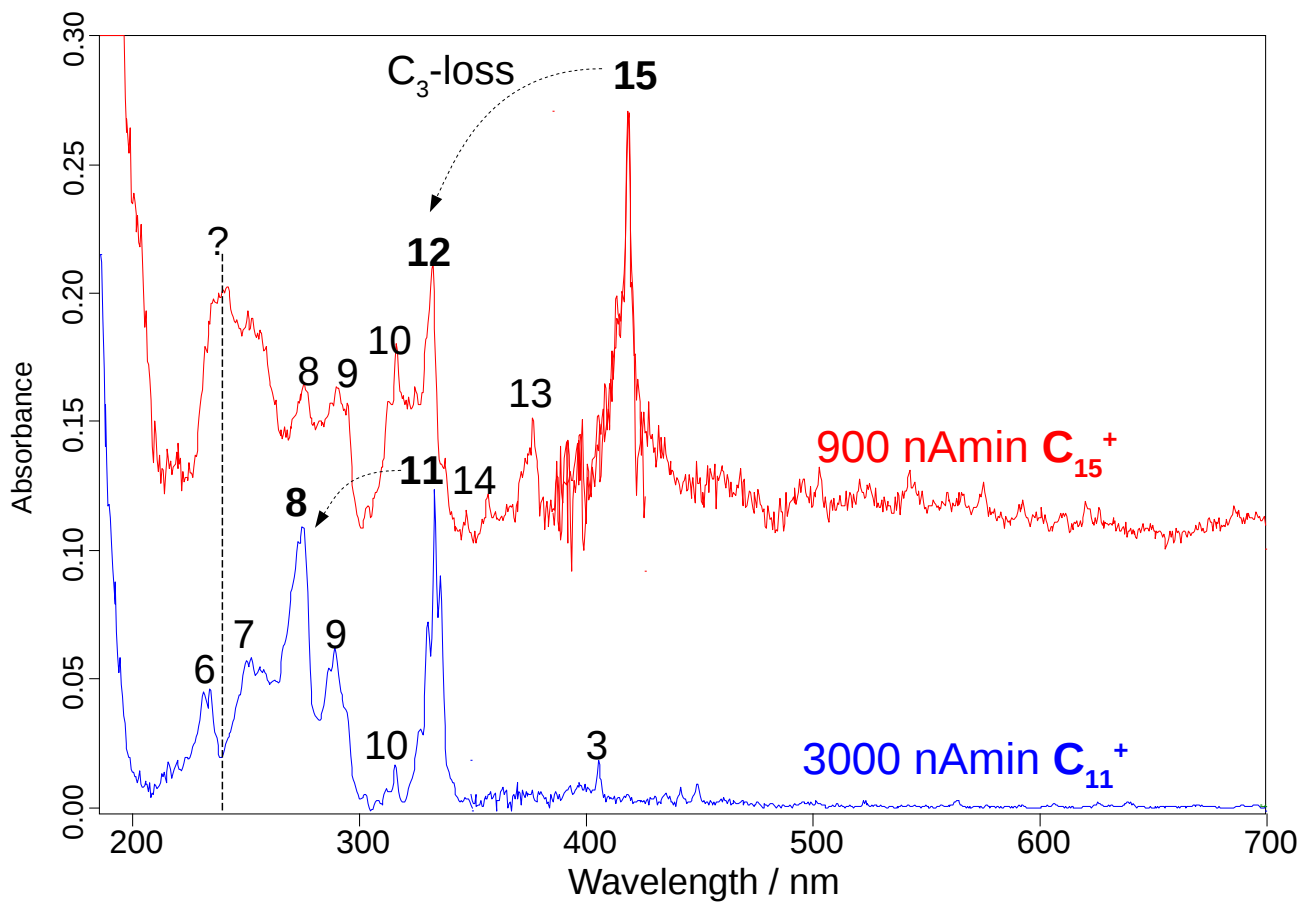}
\caption{\label{FragmentationUVVIS} UV-Vis absorption spectra in solid neon
obtained after deposition of mass-selected C$_{15}^{+}$ (red) and
C$_{11}^{+}$ (blue) at 200~eV average kinetic energy. Fragmentation
of parent chains can be observed.}
\end{figure}

Upon going to higher deposition energies of about 200~eV, we observe
(further) fragmentation of the mass-selected ions upon deposition
(Fig. \ref{FragmentationUVVIS}). Again the absorptions of these fragments
can be assigned to (smaller) linear carbon chains. We base this assignment
on previous data derived from mass-selected deposition of carbon clusters
into Ne matrixes \citep{Forney1996,Grutter1999,Wyss1999}. We find
that in our experiments the main fragmentation channel is loss of
C$_{3}$-units. Note, that C$_{10}$, C$_{12}$ and C$_{14}$ were
previously misassigned as rings \citep{Grutter1999}. This can be
concluded from the fragmentation pattern: linear C$_{15}$ should
also produce linear C$_{12}$ (Fig.\ref{FragmentationUVVIS}). Similarly,
upon depositing C$_{16}^{+}$ we observe enhanced absorption of linear
C$_{13}$ (C$_{3}$-loss) as well as absorption of linear C$_{16}^{-}$
\citep{Freivogel1995}. 

\subsection{Quantum Chemical Calculations}

\label{TC} 
\begin{figure}
\centering \includegraphics[width=0.95\textwidth]{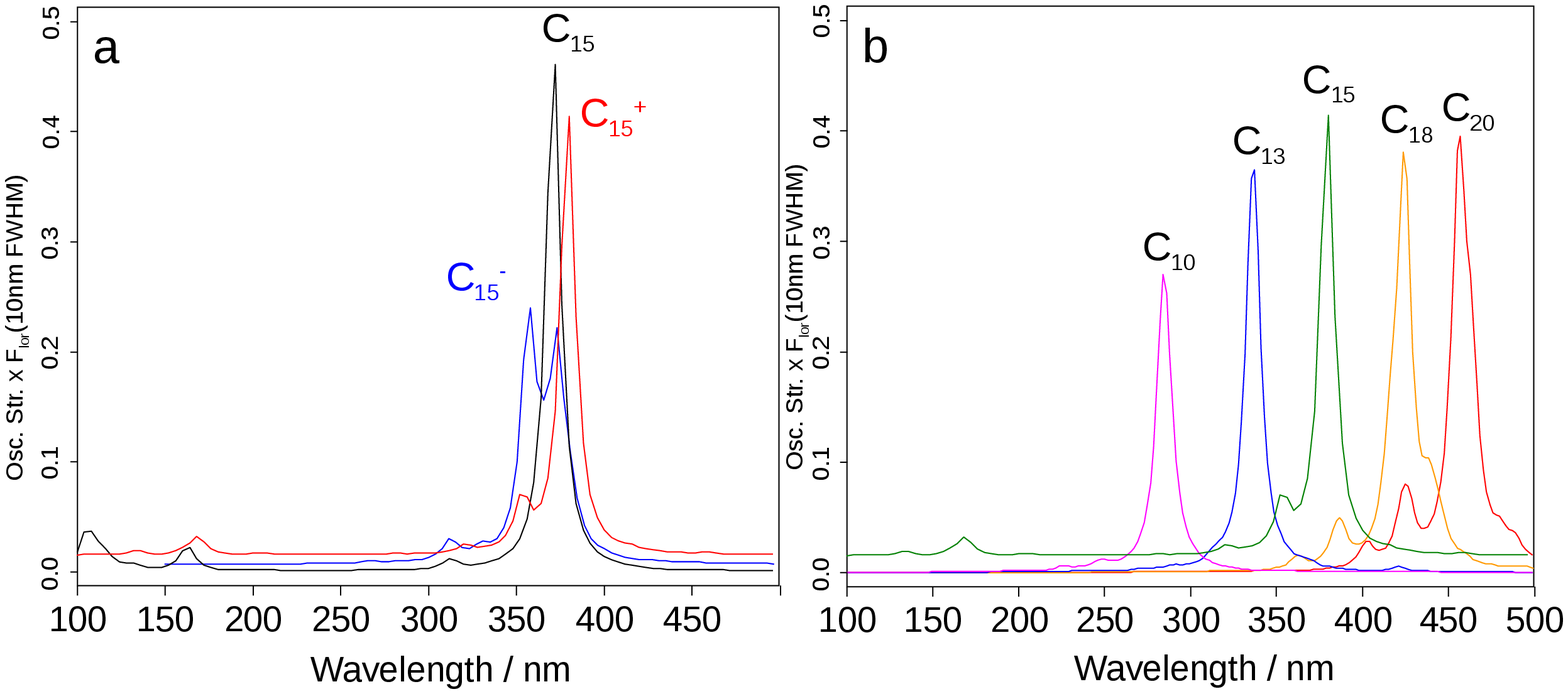}
\caption{\label{TDDFTchains} \textbf{a:} The strongest electronic absorptions
of linear C$_{15}$ and its singly charged ions according to TDDFT
(BP86/def-SV(P)) prediction, unscaled. \textbf{b:} Predictions of
the strongest electronic absorptions for a range of long neutral long
carbon chains (TDDFT (BP86/def-SV(P)), unscaled). Calculated lines
are broadened by multiplication with a 10nm-FWHM Lorentzian.}
\end{figure}

\begin{figure}
\centering \includegraphics[width=0.95\textwidth]{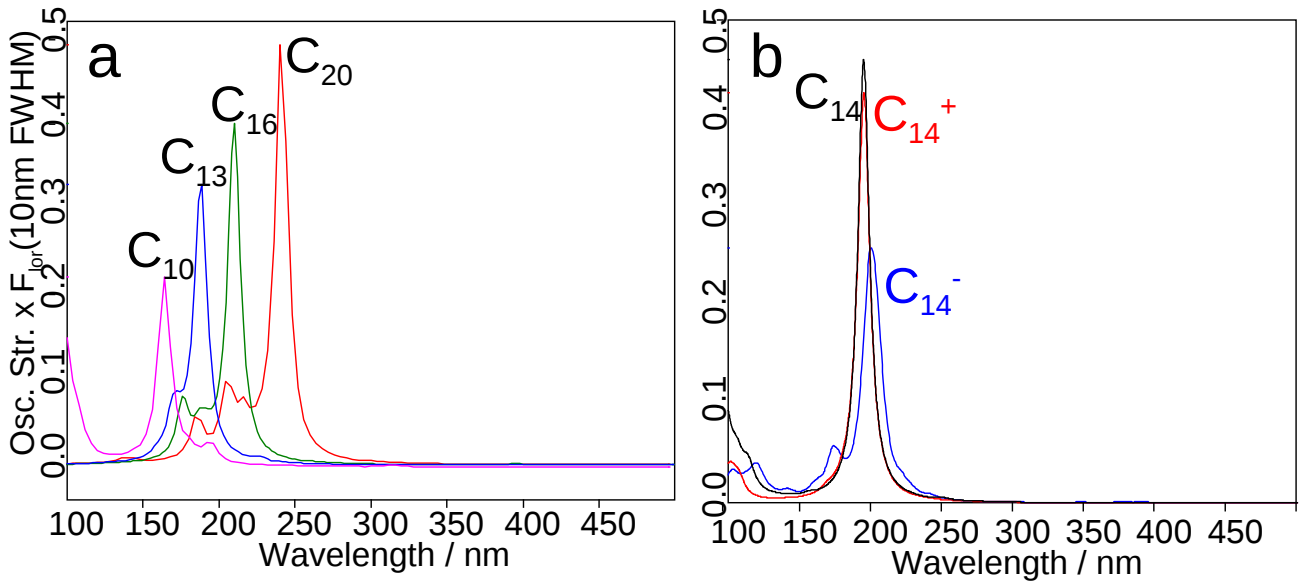}
\caption{\label{TDDFTrings} \textbf{a:} The strongest absorptions of neutral
monocyclic carbon rings as calculated by TDDFT (BP86/def-SV(P)), unscaled.
\textbf{b:} The strongest electronic absorption of C$_{14}$-ring
and its ions. TDDFT (BP86/def-SV(P)), unscaled. Calculated lines are
broadened by multiplication with a 10nm-FWHM Lorentzian.}
\end{figure}

Although, small carbon clusters have been treated theoretically many
times (\citet{VanOrden1998,Abinitio2012} and references therein),
we present own TDDFT calculations for chain and ring isomers here
with an emphasis on the wavelength ranges containing the strongest
electronic absorptions. Our results are generally consistent with
previous calculations at the same level of theory where available
(for the spectral range and cluster sizes of interest here). For all
charge states considered, ground state geometries were optimized at
the RI-DFT BP86/def2-SV(P) level of theory in $C_{1}$ and higher
symmetries, then harmonic analysis was done to check that a real ground
state was obtained. After this, energies of allowed vertical transitions
were obtained by TDDFT \citep{Turbomole}. According to our TDDFT
estimations, the most intense absorptions of linear C$_{n}^{+/-/0}$
are concentrated in comparatively narrow wavelength ranges which depend
strongly on chain length (Fig. \ref{TDDFTchains}a). These wavelength
ranges shift to the red with increasing number of carbon atoms (Fig.~\ref{TDDFTchains}b).
The strongest absorptions of mono-cyclic carbon rings according to
TDDFT are close to 200 nm (Fig.~\ref{TDDFTrings}a) --
the vacuum-UV boundary. Calculated oscillator strengths are comparable
to the strongest chain absorptions -- for the same cluster
nuclearity. There is no experimental data so far concerning these
strongest ring absorptions. We expect that they could be recorded
with a VUV-spectrometer, assuming sufficient quantity of the species
can be accumulated. Similar to linear structures, the strongest absorptions
of the ionized rings are predicted to be close to the absorptions
of the corresponding neutral ring species (Fig. \ref{TDDFTrings}b).
Note, that while ring species also have DIBs-relevant absorptions
in the visible or near-IR ranges they are however, much weaker. 

To rationalize our observation that fullerenes can be fragmented to
linear chains we have also carried out molecular dynamics (MD) simulations
of C$_{60}$ fragmentation based on the semi-empirical PM7 level of
theory \citep{MOPAC}. The initial kinetic energy of C$_{60}$-atoms
was varied from 100~eV to 500~eV and the energy dissipation time
was varied from 10~fs to 20~ps. Only a few types of different structures
result: linear chains (most abundant), monocyclic rings and chain-ring
hybrid structures. Example figures of two typical MD-simulations can
be found in the Appendix (Figs.~\ref{md150eV},\ref{md200eV}). Qualitatively similar results
were also reported previously using lower level MD simulations \citep{Kim1994}.

\subsection{Vibrational Spectroscopy}

\begin{figure}
\centering \includegraphics[width=0.5\textwidth]{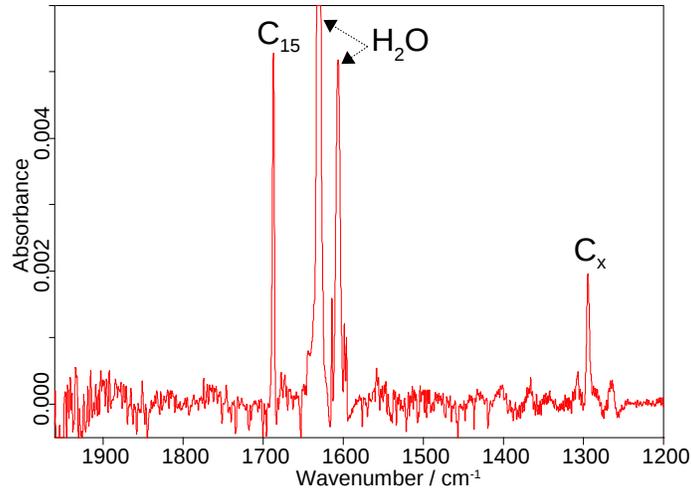}
\caption{\label{C15IR} IR absorption spectrum of C$_{15}$ in Ne at 5K, obtained
after deposition of ca. 1000 nAmin of C$_{15}^{+}$/C$_{60}^{4+}$
at about 100~eV kinetic energy. The line at 1293~cm$^{-1}$ is an
unidentified carbon molecule.}
\end{figure}

\begin{figure}
\centering \includegraphics[width=0.5\textwidth]{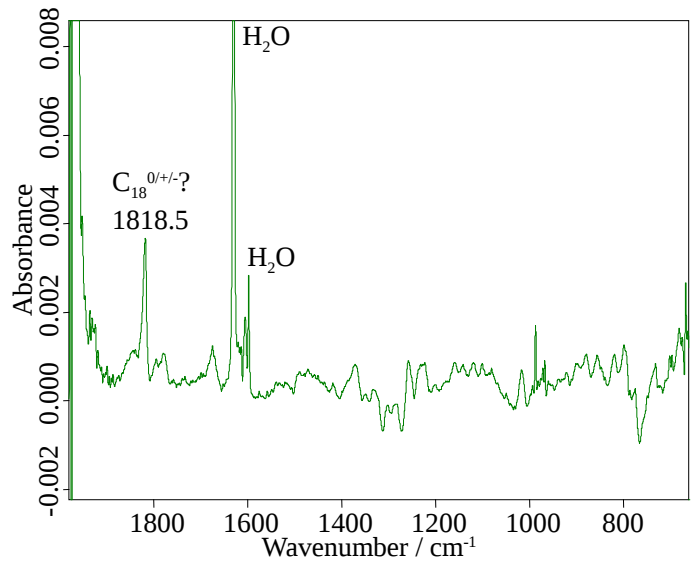}
\caption{\label{C18IR} IR absorption spectrum, obtained after deposition of
ca. 1800 nAmin of C$_{18}^{+}$/C$_{54}^{3+}$ at about 120~eV kinetic
energy in Ne at 5K. The baseline is distorted by matrix interferences.}
\end{figure}

Fig. \ref{C15IR} shows an IR absorption spectrum, which corresponds
to the UV-NIR spectrum of Fig. \ref{C15VISNIR}, i.e. after C$_{15}^{+}$
deposition under low energy impact conditions. Similar to the electronic
absorption spectrum, the spectrum looks quite clean with a very prominent
C$_{15}$ IR absorption. The IR absorption of C$_{15}$ is known from
previous experiments with a mixture of neutral matrix-isolated carbon
clusters (prepared by aggregation in matrix) \citep{Strelnikov2005}.
There is an unidentified absorption of a neutral carbon cluster C$_{x}$,
which was also present as a weak absorption line in the former IR
spectra \citep{Strelnikov2005}. This absorption does not correlate
with the IR absorption of neutral linear C$_{15}$ and remains to
be identified. Fig. \ref{C18IR} shows an IR absorption spectrum,
corresponding to the UV-NIR measurement presented in Fig. \ref{C18VISNIR}, i.e. for the matrix sample prepared by deposition of C$_{18}^{+}$/C$_{54}^{3+}$.
The absorption at 1818.5~$cm^{-1}$ belongs to a C$_{18}$-derived species.
Determination of its charge state requires further experiments with
electron scavengers. 

\section{Discussion and Outlook}

Our measurements imply that fragmentation of fullerenes in Space by
electron impact (and conceivably also by interaction with cosmic rays)
would lead to the formation of strongly absorbing linear carbon chains.
In contrast, absorption of a single photon with energy below 13.6~eV
(hydrogen cut-off) is not expected to be relevant for carbon-chain
formation as the excitation energy is insufficient for multifragmentation.
The wavelength region, within which DIBs are observed, starts at about
400~nm and extends to the near-IR \citep{Hobbs2008,Hamano2016}.
Our measurements on C$_{15}$ and C$_{18}$ containing matrixes and
the associated TDDFT calculations imply that the strongest absorptions
of linear C$_{2n+1}^{+/-/0}$ ($n>6$) and C$_{2n}^{+/-/0}$ ($n>8$)
can in fact fall in the DIBs region. The positions of the gas-phase
absorptions of neutral and ionized carbon chains are close to their
absorptions in Ne-matrix. This is known from experiments of the Maier
group on smaller carbon clusters \citep{Boguslavskiy2006,Zack2014}.
At the current stage of our experiments, absolute absorption cross-sections
are difficult to estimate, because of the overlap of C$_{n}^{+}$
with C$_{4n}^{4+}$/C$_{3n}^{3+}$/C$_{2n}^{2+}$. Furthermore, additional
experiments are needed to distinguish absorptions of C$_{n}^{+/-/0}$
in different charge states. Nevertheless, from the fact that one can
easily observe optical absorptions of C$_{15}$ and C$_{18}$ despite
their low ion currents (below 1~nA vs. ca. 100~nA for C$_{60}^{+}$),
one can deduce that the long carbon chains absorb at least one order
of magnitude more strongly than C$_{60}^{+}$ in the NIR. Theoretical
(TDDFT) oscillator strengths of the strongest C$_{n}^{+/0}$ linear
chain absorptions approximately scale as $n/2$. For any linear C$_{n}$
this is already considerably more than that of the C$_{60}^{+}$ NIR
absorption $^{2}$E$_{1g}\leftarrow$X$^{2}$A$_{1u}$ ($f_{theoretical}=0.04$,
$f_{experimental}=0.05\pm0.02$ -- integrated over all
NIR C$_{60}^{+}$ vibronic bands \citep{Depo2C60AA,C60cC70cMaierGasPhase}).
Consequently, it seems plausible that C$_{60}^{+}$ DIBs would be
accompanied by measurable absorptions due to chain-like fragments.
To prove this, we suggest that future gas-phase measurements (first
in the laboratory and then in Space) should concentrate efforts on
the chain species C$_{2n+1}^{+/-/0}$ ($n>6$) and C$_{2n}^{+/-/0}$
($n>8$). Carbon chains are known to be very reactive and in Space
might eventually become terminated by abundant heteroelements such
as H, (C), N, O. From matrix measurements, one knows that oxides of long
carbon chains have absorptions quite close to their C$_{n}$ precursors
\citep{Strelnikov2007}. Such carbon cluster oxides would therefore
also be good candidates for future gas-phase measurements.

\section{Conclusion}

Fragmentation of fullerenes upon electron impact (and conceivably
also by collision with other energetic particles) can lead to the
formation of long carbon chains with up to at least 18 carbon atoms.
The strongest absorptions of linear C$_{2n+1}^{+/-/0}$ ($n>6$) and
C$_{2n}^{+/-/0}$ ($n>8$) fall in the DIBs relevant region. Given
recent advances in the He-tagging technique \citep{MaierDIBs2015,Roithova2016}
which allows facile one-photon absorption spectroscopy of cold molecular
cations in gas phase and also the relative ease with which long cationic
carbon chains can be generated from fullerenes, we expect that corresponding
gas-phase laboratory data will now be rapidly obtained. Then Douglas'
hypothesis that carbon chains may be responsible for some of the DIBs
could finally be confirmed or disproved.

\acknowledgments This work was supported by the Deutsche Forschungsgemeinschaft
(KA 972/10-1). We also acknowledge support by KIT and Land Baden-W\"{u}rttemberg.
\newpage
\section{Appendix: MD Simulations.}
\begin{figure}[h]
\centering \includegraphics[width=0.4\textwidth]{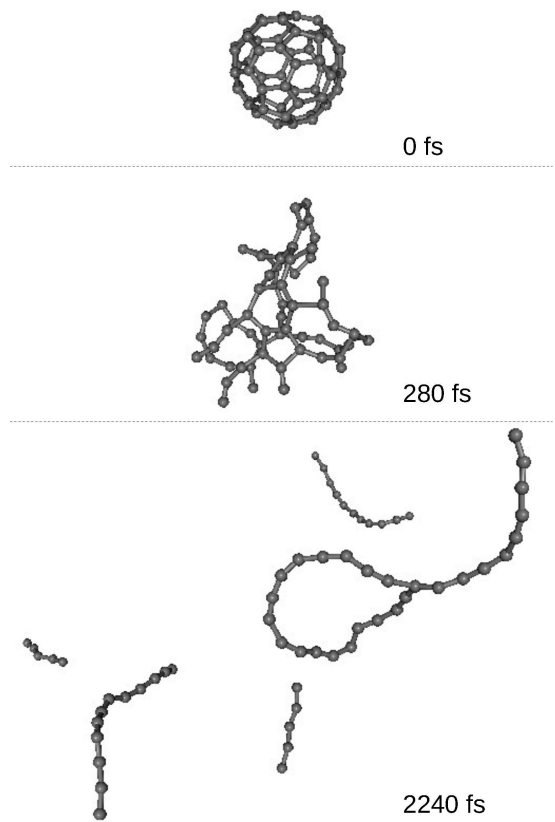}
\caption{\label{md150eV}Selected frames of the MD-simulation of C$_{60}$ fragmentation. (MOPAC, PM7 DRC, 150~eV initial energy, 1 ps energy dissipation half-life). The corresponding animation can be viewed online.}
\end{figure}
\begin{figure}[h]
\centering \includegraphics[width=0.4\textwidth]{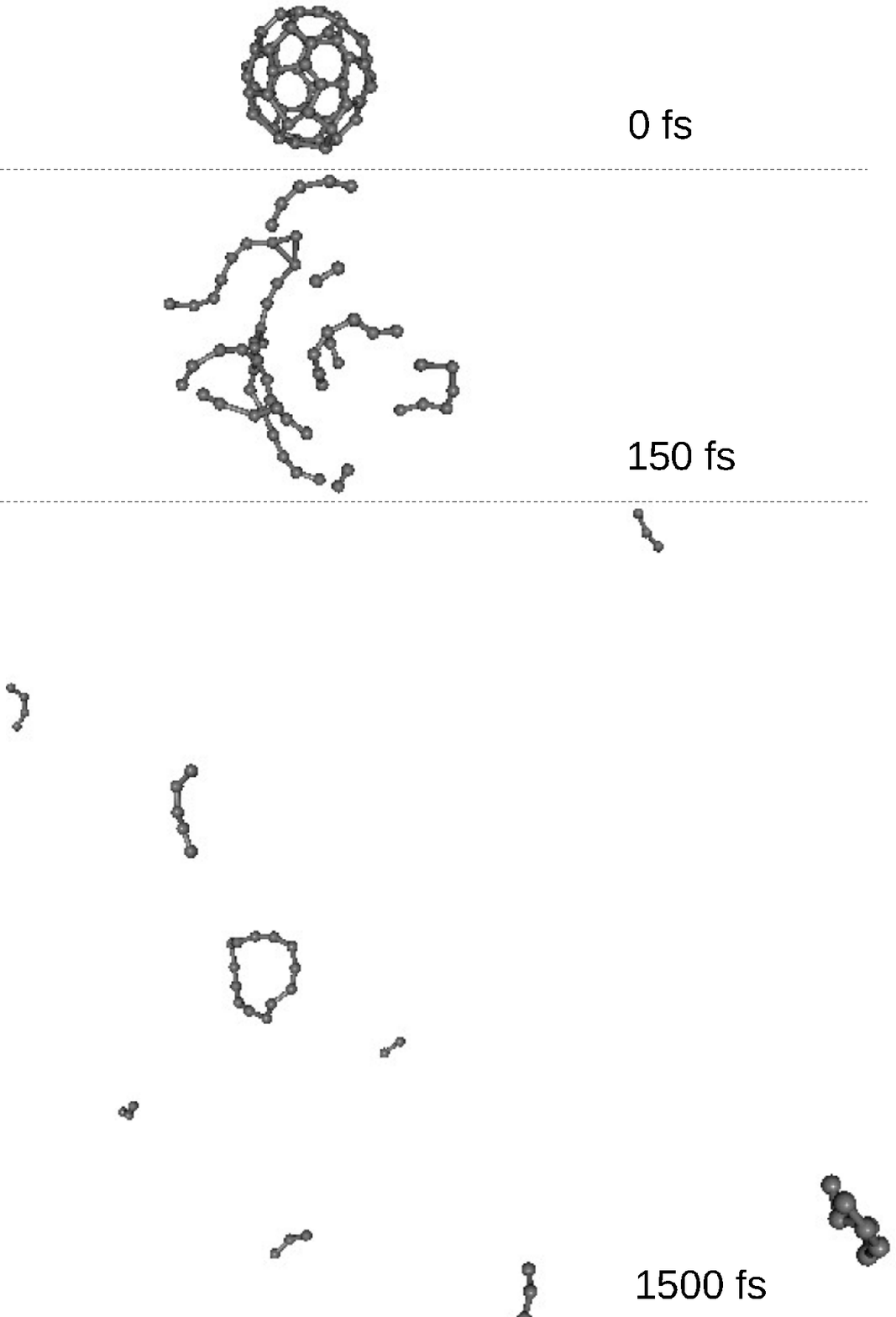}
\caption{\label{md200eV}Selected frames of the MD-simulation of C$_{60}$ fragmentation. (MOPAC, PM7 DRC, 200~eV initial energy, 4 ps energy dissipation half-life). The corresponding animation can be viewed online.}
\end{figure}

\end{document}